\documentclass[floatfix,twocolumn,pre,showpacs]{revtex4}

\usepackage{graphicx}

\usepackage{subfigure}

\usepackage{amssymb,amsmath}

\begin{document}

\title{Modeling microscopic swimmers at low Reynolds number}

\author {David J. Earl}

\affiliation{Dept. of Chemistry, University of Pittsburgh, 219 Parkman Avenue, Pittsburgh, PA 15260.}

\author {C. M. Pooley, J. F. Ryder, Irene Bredberg, and J.M. Yeomans}

\affiliation{Rudolf Peierls Centre for Theoretical Physics, 1 Keble Road, Oxford, OX1 %%@
3NP, United Kingdom.}
\date{\today}

%--------------------------------------------------------------------------

\begin{abstract}
We employ three numerical methods to explore the motion of low Reynolds number swimmers, modeling the hydrodynamic interactions by means of the Oseen tensor approximation, lattice Boltzmann simulations and multiparticle collision dynamics. By applying the methods to a three bead linear swimmer, for which exact results are known, we are able to compare and assess the effectiveness of the different approaches. We then propose a new class of low Reynolds number swimmers, generalized three bead swimmers that can change both the length of their arms and the angle between them. Hence we suggest a design for a microstructure capable of moving in three dimensions. We discuss multiple bead, linear microstructures and show that they are highly efficient swimmers. We then turn to consider the swimming motion of elastic filaments. Using multiparticle collision dynamics we show that a driven filament behaves in a qualitatively similar way to the micron-scale swimming device recently demonstrated by Dreyfus {\em et al.} \cite{DB05a}.
\end{abstract}

\maketitle

\section{Introduction}

Microscopic and mesoscopic organisms such as bacteria operate at length scales where %%@
swimming motion takes place at very low Reynolds number \cite{N2}. In his `scallop' theorem of %%@
microscopic swimming Purcell argued that swimming strategies can only be successful in %%@
this regime if they involve a cyclic and non time reversible motion
\cite{P77,SW89}. 
%This idea was put on a more rigourous footing by Shapere and Wilczek
%\cite{SW89}. 
The driving of 
helically shaped bacterial flagella by a reversible rotary engine and the beating motion  %%@
of elastic rod-like flagella utilized by eukaryotic cells are examples
of biological mechanisms which  %%@
break time reversible invariance, thus allowing microscopic organisms to move in a  %%@
controlled fashion. In an exciting recent development Dreyfus {\em et al.} \cite{DB05a} %%@
have demonstrated for the first time the controlled swimming motion of a fabricated, %%@
micrometer size device. 

Several authors have described models of swimmers at low Reynolds number. Simple models %%@
which comprise linked spheres or connected rods that move by changing the distances or directions %%@
between the components are considered in \cite{P77,BK03,NG04,AK05,DB05b}.
% check overlap with our work eg for rotating Purcall swimmer
For one dimensional motion analytic results for the swimming velocity and efficiency can %%@
be obtained. 
Felderhof \cite{F06} used the Oseen tensor formalism to model microscopic swimmers using a 
bead spring model. Gauger and Stark \cite{GS06} used a similar method to model the experimental 
elastic filament of Dreyfus {\em et al.} \cite{DB05a}. Both of these approaches are distinct from the 
one we use, in that the actuation of the beads is described in terms of forces, 
whereas we define swimming in terms of a predefined shape change.
Propulsion by a non time reversible pattern of surface distortions is addressed in %%@
 \cite{SS96,Y06a,PB06}. Another possible swimming mechanism, mediated by an asymmetric %%@
distribution of reaction products, is proposed in \cite{GL05}. 

Increasingly, quantitative experiments on bacterial dynamics are appearing in the %%@
literature. Transient collective motion has been observed in collections of swimming %%@
cells \cite{DC04}. Bacteria near solid boundaries have been shown to swim in %%@
circles \cite{LD06} and those near an obstacle to reverse their swimming %%@
direction \cite{CD06}. Experiments have been performed to determine the dependence of the %%@
chemotactic response of \emph{Dictyostelium discoideum} cells swimming
in a concentration %%@
gradient \cite{B06}.

Although simple, analytical models can provide considerable help in understanding these %%@
results there are many new features inherent or accessible in real biological systems %%@
that remain to be explored. These include more complicated swimming mechanisms, interactions between %%@
densely packed swimmers and the effect of boundaries and obstacles. There will %%@
increasingly be a need to develop numerical methods to probe the behavior of more %%@
complicated structures and situations where analytic approaches become intractable. 

In numerical approaches published so far Hernandez-Ortiz {\em et al.} have considered
the swimming motion of a collection of force dipoles \cite{HS05}. They
observe the large scale coherent vortex motion that has been seen in
experiments. 
Ramachandran {\em et al.} have described swimmers, modeled as force dipoles,
interacting with a fluid described by a lattice Boltzmann 
algorithm \cite{RK06}. 
Work on swimmers propelled by a flexible filament modeled
hydrodynamics through an anisotropic friction coefficient \cite{L03, LC03}.

Our first aim in this paper is to explore new ways in which simulation
methods can be applied to the motion of swimmers in a low Reynolds
number solvent. We model the hydrodynamic interactions by using
the Oseen tensor approximation \cite{Oseenref}, a lattice 
Boltzmann algorithm \cite{S01,Y06} and multiparticle collision
dynamics \cite{MK99}. The approaches are validated by solving the
equations of motion for a linear three bead swimmer where an
analytic solution is available for comparison \cite{NG04}. We discuss
the relative merits and demerits of the three approaches.

Secondly, we extend the linear model to more general three bead microstructures. We use an %%@
Oseen tensor approach to demonstrate that they can move in a controlled fashion in three %%@
dimensions by changing both the length of and the angle between their arms and we discuss the %%@
efficiencies of the various swimming modes. We also show that multibead linear %%@
structures are highly efficient swimmers. 

We then consider the swimming motion of driven elastic filaments. Our model, solved using %%@
multiparticle collision dynamics, mirrors the behavior of the swimming device introduced %%@
by Dreyfus {\em et al.} \cite{DB05a}. In the conclusion, we consider future directions in %%@
which the modeling approaches might be useful in understanding the swimming of bacteria %%@
and of fabricated microstructures.

\section{Numerical Approaches}
\label{numsec}

The swimmers we consider are composed of $N$ spheres, 
of fixed radius $R$ and with 
positions given by ${\bf r}_{i}$, where 
$i = 1...N$. The spheres are linked 
by rods that are sufficiently thin to neglect any
hydrodynamic effect. Internal forces and torques act to change 
the lengths and/or angles between the rods, causing the swimmer 
to change shape. These shape changes, when coupled to the fluid, 
lead to directed motion. We now describe three different numerical 
approaches used to simulate the fluid.

\subsection{Oseen tensor}

\label{OseenT}

The Oseen tensor allows us to consider the hydrodynamic interaction, in the 
limit of zero Reynolds number, between spheres that are spaced far 
apart ({\em i.e.} at distances significantly larger than their radii, $R$) \cite{Oseenref}. 
A sphere pushed by a force will move, and so set up a flow 
field. Any surrounding spheres will be advected 
with the resulting local fluid velocity. 
%Thus, a force on one sphere will produce a velocity on another. 
Furthermore, since the Reynolds number is very low, 
the time taken to set up the flow fields is much smaller than 
that needed for a given sphere to move a significant fraction of 
$R$ \cite{N1}. Therefore, the hydrodynamic 
interaction can be approximated as instantaneous. 
Since the Stokes equation (the Navier-Stokes equation without 
the inertial term, as is appropriate in the low Reynolds number limit)
is linear, the velocity fields produced by each of the spheres simply add up. 
This allows us to write
\begin{equation}
%{\bf v}_i^\alpha = \sum_{\beta = 1}^{M} \sum_{j = 1}^N {\bf H}_{ij}^{\alpha \beta} {\bf F}_j^\beta,
{\bf v}_i = \sum_{j = 1}^N {\bf H}_{ij} {\bf F}_j
\label{oseen}
\end{equation}
where ${\bf v}_i$ is the velocity of sphere $i$ and ${\bf F}_j$ is the
force on sphere $j$. The Oseen tensor ${\bf H}_{ij}$ is \cite{Oseenref} 
\begin{equation}
{\bf H}_{ij} =
\begin{cases}
\frac{{\bf I}}{6\pi\eta R},  &\text{if $i = j$}, \\
\frac{1}{8\pi\eta | r_{ij} |} \left( {\bf I} +
  \frac{{\bf r}_{ij} {\bf r}_{ij}}{ |
    r_{ij} |^2} \right) &\text{otherwise},
\end{cases}
\label{oseent}
\end{equation} 
where $\eta$ is the fluid viscosity, ${\bf I}$ is the identity matrix,
and ${\bf r}_{ij} = {\bf r}_j - {\bf r}_i$
is the vector between spheres $i$ and $j$.
For a swimmer, the forces ${\bf F}_i$ are not external, but rather internal 
forces mediated through the links that connect the spheres.
They are subject to the constraints 
\begin{equation}
%\sum_{i=1}^N {\bf F}_i^\alpha = 0, \quad  \sum_{i=1}^{N} {\bf F}_i^\alpha \times {\bf r}_i^\alpha = 0,
\sum_{i=1}^N {\bf F}_i = 0, \quad  \sum_{i=1}^{N} {\bf F}_i \times {\bf r}_i = 0
\label{constraint}
\end{equation}
which state that no external forces or torques act on the swimmer. 

\begin{figure}
\begin{center}
\includegraphics[width=3in]{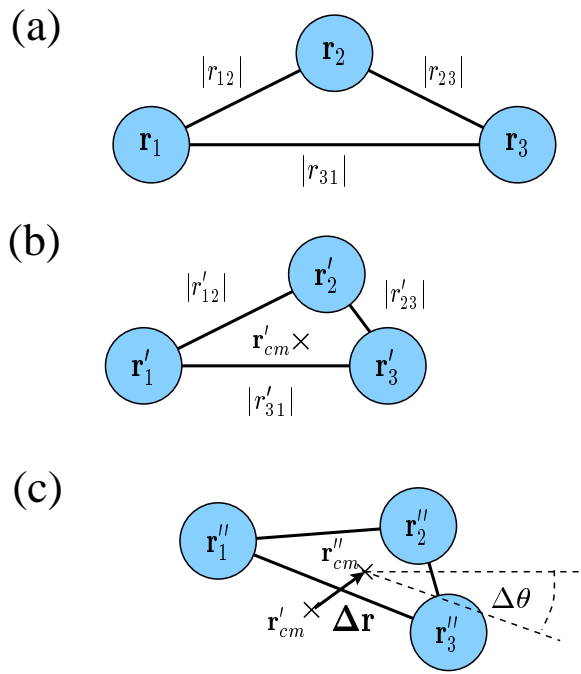}\\
\end{center}
\caption{
(a) The swimmer at time $t$. 
(b) The swimmer {\it shape} defined at time $t+\delta t$. The lengths
of the two links connecting sphere $3$ have decreased.
(c) The shape at $t+\delta t$ is translated by a vector 
$\Delta {\bf r}$ 
and rotated by an angle $\Delta \theta$ around its center of mass. 
These operations preserve the link lengths, thus leaving the shape 
unchanged. The parameters $\Delta {\bf r}$ and $\Delta \theta$ are 
chosen to improve upon the accuracy of the constraints in equation 
(\ref{constraint}). This procedure is performed iteratively until 
the necessary accuracy is achieved.  
}
\label{figoseen}
\end{figure}

The swimming motion is defined as a periodic {\it shape} change
and, from this information, the algorithm must determine the 
trajectory of the swimmer through the fluid. 
To illustrate how this works we begin by considering a 
swimmer whose motion is confined to a two dimensional plane. 
Figure \ref{figoseen} shows the procedure for the case $N = 3$. 
At a given time $t$, the position of the spheres ${\bf r}_i$ are
known, as shown in figure \ref{figoseen}(a). 
The new shape of the swimmer at the next time step 
$t+\delta t$ is shown in figure \ref{figoseen}(b). 
We have chosen, for illustrative purposes, a swimming step where
the lengths of the two links connecting sphere 3 have decreased.

The shape of the swimmer in figure \ref{figoseen}(b) is defined 
through the three quantities $|r_{12}^\prime|$, $|r_{23}^\prime|$, 
and $|r_{31}^\prime|$. However, this information does not determine 
the absolute positions of the spheres.
%${\bf r}_1^\prime$, ${\bf r}_2^\prime$, and ${\bf r}_3^\prime$. 
To find these, it is necessary to enforce the 
conservation conditions stated in equation (\ref{constraint}). This 
is performed in the following iterative manner. 

We take the first approximation for ${\bf r}_i(t+\delta t)$ to be 
${\bf r}_i^\prime$. This does not, in general, obey 
equation (\ref{constraint}), as will be apparent below. Our aim is to
move the spheres in such a way as to successively improve the 
accuracy of equation (\ref{constraint}). To do this, we consider translating 
the swimmer by a vector $\Delta {\bf r} = (\Delta x, \Delta y)^T$, 
and rotating it about its center of mass by an angle $\Delta \theta$, 
as illustrated in Figure \ref{figoseen}(c). Note that these operations 
do not change the shape of the swimmer.
In doing this, we introduce new position coordinates 
${\bf r}_i^{\prime \prime}$ defined by 
\begin{eqnarray}
{\bf r}_i^{\prime \prime} = {\bf r}^\prime_{cm} + {\bf \Delta r} + {\bf R} \left( {\bf %%@
r}^\prime_i - {\bf r}^\prime_{cm} \right) 
\label{better}
\end{eqnarray}
where ${\bf R}$ is a clockwise rotation matrix around an angle 
$\Delta \theta$ and 
${\bf r}^\prime_{cm} = \sum_{i=1}^N {\bf r}_i^\prime/N$ is the center
of mass position of the swimmer.
The unknown displacement ${\bf \Delta r}$ and angle $\Delta \theta$ 
are chosen to improve upon the accuracy of the constraints in equation
(\ref{constraint}). To calculate them we note that
the velocity of sphere $i$ can be related to its displacement over 
time $\delta t$ by
\begin{eqnarray}
{\bf v}_i &=& \frac{{\bf r}_i^{\prime \prime} -  {\bf r}_i}{\delta t} \nonumber\\ 
&\approx& \frac{{\bf r}_{i}^\prime -  {\bf r}_i + {\bf \Delta r} + \Delta \theta |{ %%@
r}_i^\prime - { r}_{cm}^\prime| \hat{\bf r}_i^\theta}{\delta t}. \label{v}
\end{eqnarray}
The unit vector $\hat{\bf r}_i^\theta$ lies in the direction a given point moves 
under an infinitesimal rotation (which can be calculated by
rotating the vector ${\bf r}^\prime_i - {\bf r}^\prime_{cm}$ by 
$90^o$ clockwise and normalizing it).

Substituting expression (\ref{v}) into equation (\ref{oseen}), and 
numerically inverting the resulting matrix equation using Gaussian 
elimination, we obtain an expression for the forces acting on 
each sphere of the form
\begin{eqnarray}        
{\bf F}_i = {\bf a}_i + {\bf b}_i \Delta x + {\bf c}_i \Delta y + {\bf d}_i \Delta \theta  
\label{F}
\end{eqnarray}
where ${\bf a}_i$, ${\bf b}_i$, ${\bf c}_i$, and ${\bf d}_i$ 
are constant vectors.
Using this, the three constraints in equation (\ref{constraint})
can, after some rearranging, be written
\begin{eqnarray}
{\bf A} 
\left(
\begin{array}{c}
\Delta x \\
\Delta y \\
\Delta \theta
\end{array}
\right)
= 
\left(
\begin{array}{c}
e_1 \\
e_2 \\
e_3 
\end{array}
\right),
\label{A}
\end{eqnarray}
where the matrix ${\bf A}$ and the column vector ${\bf e}$ are 
constants. Finally, this matrix equation can be inverted to obtain 
the values of $\Delta x$, $\Delta y$, and $\Delta \theta$ which are then
used in equation (\ref{better}) to obtain a better approximation, 
${\bf r}^{\prime\prime}$, to the sphere positions at time
$t+\delta t$.
%This then replaces the old ${\bf r}^{\prime}$, and the 
%procedure is repeated. After successive iterations we obtain 
Repeating the procedure gives rapid 
convergence to the correct solution. Typically a single iteration 
improves the accuracy of equation (\ref{constraint}) by a factor 
of $\sim 10$.

The approach is easily generalized to a swimmer moving in three
dimensions. In this case there are three 
displacement parameters $\Delta x$, $\Delta y$, $\Delta z$, and three 
rotation parameters $\Delta \theta_{xy}$ $\Delta \theta_{yz}$, 
$\Delta \theta_{zx}$ (where, for example, $\Delta \theta_{xy}$ is a rotation in the 
$x$-$y$ plane). These six unknowns can be determined using the six 
constraints in equation (\ref{constraint}). The method is the same as 
above, except equations (\ref{better}-\ref{F}) now contain the 
corresponding extra terms, and ${\bf A}$ in equation (\ref{A}) 
becomes a six by six matrix.
 
The computation for one time step scales as $N^3$. For low to moderate values of $N$ the %%@
method is very fast. However, for $N \gtrsim 100$, it quickly becomes very computationally %%@
intensive.

\subsection{Lattice-Boltzmann}

\label{latboltsec}

The lattice Boltzmann algorithm is now a widely used mesoscopic
modeling technique for simulating the behavior of complex fluids \cite{S01,Y06}.
The method consists of an evolution equation for a mass density 
distribution function $f_k({\bf s},t)$, which can be considered as a 
simplified, discretized version of Boltzmann's transport equation. 
The distribution function is defined at positions, ${\bf s}$, which
lie on a cubic lattice 
with a distance $\delta s$ between nearest neighbor points. 
Its value is updated simultaneously and discretely in 
time, with time step $\delta t$. We define $c = \delta s / \delta t$.
The subscript $k$ denotes a particular velocity 
direction ${\bf e}_k$.
The velocity vectors ${\bf e}_k$ 
must be chosen such that ${\bf e}_k \delta t$ lies between lattice sites.
In this study all simulations are performed in three dimension using a
$15$ velocity model. This has a zero velocity vector 
${\bf e}_0 = (0,0,0)$, six nearest neighbor velocity vectors 
${\bf e}_{1-6}$ in the directions $(\pm c,0,0)$, $(0,\pm c,0)$, and 
$(0,0,\pm c)$, and eight velocity vectors ${\bf e}_{7-14}$ in the 
diagonal directions $(\pm c, \pm c, \pm c)$.
From $f_k$ we can calculate the mass density $\rho$ and momentum 
density $\rho {\bf u}$:
\begin{eqnarray}
\rho=\sum_k f_k,&\ \ \ \ \ \ &\rho u_{\alpha}=\sum_k f_k e_{k \alpha}.
\label{5.3}
\end{eqnarray}

Evolution in time is given by
\begin{eqnarray}
 f_k({\bf s} + {\bf e}_k \delta t , t+\delta t) = f_k({\bf s}, t) - \frac{1}{\tau} \Bigl[ %%@
f_k({\bf s}) - f_k^{eq}({\bf s}) \Bigr]
\label{latbolt}
\end{eqnarray}
where we use the Bhatnagar-Gross-Krook approximation, which uses a 
single parameter $\tau$ to determine the 
rate of relaxation toward equilibrium.
A suitable choice for the equilibrium distribution function is 
\begin{eqnarray}
f^{eq}_k &=& \rho w_k \left( 1 + \frac{3 e_{k\alpha} u_\alpha}{c^2} + \frac{9 %%@
(e_{k\alpha} u_\alpha)^2}{2c^4} - \frac{3 u^2}{2 c^2} \right)
\label{fequ}
\end{eqnarray}
where the weight factors are 
$w_0 = \frac{2}{9}$, $w_{1-6} = \frac{1}{9}$, 
and $w_{7-14} = \frac{1}{72}$. 
Note that this distribution satisfies
\begin{eqnarray}
\sum_k f_k^{eq} = \rho,&\ \ \ \ \ \ 
&\sum_k f_k^{eq}e_{k \alpha}= \rho u_{\alpha},
\label{ncons}
\end{eqnarray}
such that mass and momentum are conserved in time. This can be seen 
by summing the zeroth and first velocity moments of equation (\ref{latbolt}), and 
using the relations in (\ref{5.3}).

Applying the Chapman-Enskog expansion to the lattice Boltzmann 
equation (\ref{latbolt}) \cite{Y06} gives 
the continuity equation for the total density,
\begin{equation}
\partial_t \rho +\partial_\alpha( \rho u_\alpha) = 0
\label{conteqn}
\end{equation}
and the Navier-Stokes equation for the fluid momentum,
\begin{eqnarray}
\partial_{t}(\rho u_\alpha) +  \partial_\beta ( \rho u_\alpha u_\beta) = -  %%@
\partial_\alpha \left( \rho c_s^2 \right) + \partial_\beta \left(  \nu \rho %%@
\partial_\beta u_\alpha \right)
\label{nsfinal2}
\end{eqnarray}
where the kinematic viscosity is 
\begin{eqnarray}
\nu = \frac{(\delta s)^2}{3 \delta t} \left(\tau - \frac{1}{2} \right)
\end{eqnarray}
and the speed of sound is $c_s = c/\sqrt{3}$. At low Reynolds numbers the fluid %%@
is, to a very good approximation, incompressible, {\em i.e.} $\nabla.{\bf u} = 0$.
(In the simulations, the difference between the maximum and minimum density within 
the system was found to be no more that $0.04\%$ of the average density.)

For simplicity, we do not explicitly consider a solid-fluid interface but define 
the swimmer to comprise of the fluid region within the spheres which make up the 
swimmer \cite{N3}. The swimmer-fluid interaction is incorporated into the lattice 
Boltzmann algorithm after finding the fluid velocities in equation (8), but before 
calculating the equilibrium distributions in equation (10). This interaction is generated in three 
stages. Firstly, the total linear and angular momentum of the swimmer are calculated:
\begin{eqnarray}
{\bf P} = \sum_j \rho_j {\bf u}_j, \quad {\bf L} = \sum_j {\bf s}_j \times \rho_j {\bf u}_j,
\end{eqnarray}
where the sum $j$ runs over all the lattice sites contained within the swimmer. 
Secondly, the new positions of the spheres are calculated. This procedure is 
analogous to that described for the Oseen tensor method in Section \ref{OseenT}. In this, we know 
the positions of the spheres at time $t$ and the swimmer {\it shape} at time 
$t+\delta t$. The algorithm works out how this new shape is oriented with respect to 
the original, such that linear and angular momentum are conserved, {\em i.e.}
\begin{eqnarray}
\sum_i m_i {\bf v}_i = {\bf P}, \quad \sum_i {\bf r}_i \times m_i {\bf v}_i = {\bf L},
\end{eqnarray}
where $m_i = \sum_j \rho_j$ and 
\begin{eqnarray}
{\bf v}_i = \frac{{\bf r}_i(t+\delta t) - {\bf r}_i(t)}{\delta t}
\end{eqnarray} 
are the mass and velocity of sphere $i$, respectively. Thirdly, the 
motion of the swimmer is coupled back to the fluid. 
Lattice sites within a given sphere are set to the velocity of that sphere, 
{\em i.e.} ${\bf u}_j = {\bf v}_i, \,\,\, {\bf s}_j \in \text{Sphere}\,\, i$. These 
updated velocities are then used in calculating the equilibrium distributions 
(\ref{fequ}). Through repeated iteration of the lattice Boltzmann equation (9),
the fluid within and immediately adjacent to the spheres is
strongly coupled to move with the same velocity as the spheres, thus giving the correct boundary condition. 

To avoid unwanted lattice effects, the radius $R$ of each sphere must be 
sufficiently large to accurately resolve its shape on the cubic grid. In this study 
we choose $R=3\delta s$, such that each sphere contains approximately $113$ lattice 
sites. Note that in this procedure we neglect the effect of rotation on the spheres, assuming that all parts of a given sphere travel at the same velocity.
This assumption is justified because the hydrodynamic interactions between rotating spheres is rather weak. 
(The Oseen tensor (\ref{oseent}) decays as $r^{-1}$, whereas the flow field around a rotating sphere decays at a much faster rate of $r^{-3}$.) 
In the case of modeling more than one swimmer, it is necessary to add hard core 
potentials between spheres to prevent them from overlapping.

\subsection{Multiparticle Collision Dynamics}

\label{SRDsec}

An alternative mesoscale approach, which solves the equations
of {\em fluctuating} hydrodynamics
is the multiparticle collision dynamics (also known as stochastic rotation
dynamics) algorithm introduced by 
Malevanets and Kapral \cite{MK99}. 
The fluid is represented by a large number of point-like
particles of mass $m$, with position $\mathbf{r}_k(t)$, and 
velocity $\mathbf{v}_{k}(t)$ at time $t$, where $k$ is
the particle index. The particles move in continuous space, and
at discrete time intervals, $\delta t$. Particle positions are updated
according to

\begin{equation}
\mathbf{r}_k(t+\delta t) = \mathbf{r}_k(t) + \mathbf{v}_k(t)\delta t.
\label{eq:SRD1}
\end{equation}
At each time step the particles also undergo a multiparticle collision
that locally conserves mass, momentum, and energy. To perform the collision,
the simulation box is divided into a grid of cubic cells, with sides
of length $a$. The average number of particles per cell will be
denoted by $\gamma$. The velocities of the particles in each cell are
then rotated about the center of mass velocity of the cell,
$\mathbf{v}_{cm}$
\begin{equation}
\mathbf{v}_k(t+\delta t) = \mathbf{v}_{cm}(t) +
\mathbf{R}\left[\mathbf{v}_k(t) - \mathbf{v}_{cm}(t)\right] .
\label{eq:SRD2}
\end{equation}
$\mathbf{R}$ is a rotation matrix through a 
fixed angle, $\alpha$, about an axis that is generated randomly for each cell in the %%@
simulation at each time step. The position of the cubic grid is chosen randomly at each %%@
time step -- this leads to substantially improved Galilean invariance in the algorithm %%@
 \cite{IK01}. In the continuum limit, the multiparticle collision dynamics algorithm %%@
recovers the thermohydrodynamic equations of motion and thus acts as a Navier-Stokes %%@
solver. Conveniently, the dependence of the transport coefficients of the fluid on the %%@
simulation parameters is known analytically \cite{IK03,PY05}. Thus, it is a relatively %%@
simple task to choose values that result in a low Reynolds number fluid. 

We couple a swimmer to the multiparticle collision dynamics solvent 
by considering it to be composed of a number of particles, 
and including these solute particles in the solvent collision step 
(\ref{eq:SRD2}). In this way the swimming microstructure can exchange 
momentum with the solvent. In general, the equations of motion of
the microstructure are solved using a velocity Verlet molecular 
dynamics algorithm. Precise details of how specific swimmers
are dealt with using this approach are given in later sections
of the paper.

%The equations of motion for the microstructure
%are solved using a velocity Verlet molecular dynamics algorithm, and
%the solute exchanges momentum with the SRD solvent by including
%the solute particles in the solvent collisional step (equation \ref{eq:SRD2}).

\section{Linear Three Sphere Swimmer}
\label{1-D}

\begin{figure}
\begin{center}
\includegraphics[width=3in,height = 4.9in]{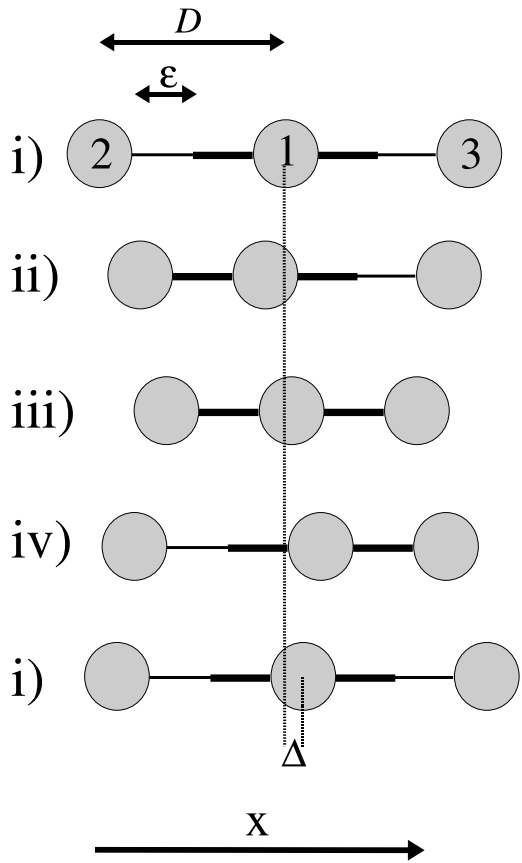}
\end{center}
\caption{The four step, cyclic motion of the linear three sphere
swimmer \cite{NG04}.}
\label{fig:gol1}
\end{figure}

Recently, Najafi and Golestanian~\cite{NG04} proposed a one-dimensional
swimmer comprising three connected spheres. Their model is perhaps the
simplest example of a controlled, cyclic motion that breaks time
reversibility. The swimmer consists of a central sphere that is
connected to two other spheres by arms that are separated by an
angle of $180^o$, are of negligible thickness, and whose
length can be changed by, for example, the action of engines located
on the central sphere. The microstructure moves by shortening and 
extending the lengths of the arms in a periodic and time irreversible 
manner, as shown in Figure \ref{fig:gol1}. The relevant parameters 
for this model
are $D$, the distance between the central sphere and an outer sphere
at the maximum arm length, $\varepsilon$, the distance the arm shortens,
$W$, the speed at which the arms change their lengths, and $R$, the 
radius of the spheres. The result of this cyclic, time
irreversible motion is a net translation of the swimmer along the
line linking the three spheres; we define $\Delta$ as the distance 
the swimmer translates in one complete cycle.

\subsection{Analytic Theory}

Because of the simplicity of the shape deformations of the linear three
sphere swimmer, it is possible to calculate analytically the total
net displacement, $\Delta$, of the swimmer during each complete cycle 
of its motion, in the limit of $\varepsilon \ll D$ and $R \ll D$.
We summarize the argument of Najafi and Golestanian \cite{NG04}
 and correct their formula for the displacement
of the swimmer, which we shall need for comparison to the numerical
results.

%Now, considering the specific case of the three linked sphere swimmer, 
%during the four steps that make up one complete cycle of motion, the
%arms of the swimmer extend and contract at the constant speed, $W$.
%As the spheres only move along the axis connecting the spheres, 
%consideration of the swimmer's motion can be reduced to a one-dimensional
%problem. Following Najafi {\em et al}.~\cite{Najafi}, 
Each of the four
steps of the stroke can, by a simple transformation, be converted into
one particular auxiliary stroke. In the auxiliary stroke one arm
has a fixed length, $\delta$, where $\delta$ is either $D$ or 
$D - \varepsilon$, and the other arm changes length from
$D$ to $D - \varepsilon$. 
%By calculating the 
%displacement of the middle sphere during the auxiliary stroke, 
%$\Delta_a(\delta)$, 
%one can determine the total displacement of the swimmer during one
%complete four stroke cycle. Taking into consideration how each
%of the four strokes is related to the auxiliary stroke, 
%Najafi and Golestanian~\cite{NG04} showed that the total
%displacement of the swimmer in one stroke, $\Delta$, is
%\begin{equation}
%\Delta = 2 \left[ \Delta_a(D) - \Delta_a(D - \varepsilon) \right].
%\label{eq:th5}
%\end{equation}
%As one complete stroke cycle is performed in a time $4 \varepsilon / W$,
%the mean swimming velocity is
%
%\begin{equation}
%U_s = \frac{\Delta}{(4 \varepsilon / W)}.
%\label{eq:th6}
%\end{equation}
%The motion of the three spheres during the auxiliary stroke can
%be modeled by considering the spheres as moving due to the forces
%exerted on them. The relationship between the velocities of each of
%the spheres is given by equation (\ref{oseen}).
%There is the requirement that the swimmer is force free (\ref{constraint})
%
%\begin{equation}
%\sum_{\text{i = 1}}^3 \mathbf{F}_i = 0
%\label{eq:th7}
%\end{equation}
%and 
We choose the {\em x}-axis to be parallel to the line linking the spheres 
and directed away from sphere $2$ (see Figure \ref{fig:gol1}). 
During the auxiliary stroke $v_1 = v_3$ and 
$W = v_2 - v_1$, and thus the velocity of the middle sphere, 
in the limit that the swimmer undergoes small deformations, is

\begin{equation}
\begin{aligned}
v_1(\delta) &\approx \frac{-W\left(H_{11}-H_{23}-H_{12}\right)}
                   {\left(3H_{11}-2\left(H_{12}+H_{13}+H_{23}\right)\right)}\\
            &\approx
            -\frac{W}{3}\left[1-\frac{R}{2(D-Wt)}+\frac{R}{\delta}
                              -\frac{R}{2(\delta+D-Wt)}\right],
\end{aligned}
\label{eq:th8}
\end{equation}
ignoring terms of order $(R/D)^2$ and greater.
The elements of the Oseen tensor for each pair of spheres
follow from equation (\ref{oseent}). 
Integrating (\ref{eq:th8}) gives the %%@
displacement over the auxiliary stroke,

\begin{equation}
\Delta_a(\delta) = \int^{\varepsilon/W}_0 v_1(\delta)\,dt.
\label{eq:th9}
\end{equation}
This can then be used to calculate the total displacement after the four step cycle, to second order in $\varepsilon / D$, as

\begin{eqnarray}
\Delta  &=& 2 \left[ \Delta_a(D) - \Delta_a(D - \varepsilon) \right] \nonumber \\
        &=& \frac{7}{12} R 
      \left[
      \left(\frac{\varepsilon}{D}\right)^2 +
      \left(\frac{\varepsilon}{D}\right)^3
      \right],
\label{eq:th13}
\end{eqnarray}
%Substituting for $\Delta$ in equation (\ref{eq:th6}), 
%the mean swimming velocity is

%\begin{equation}
%U_s = \frac{WD}{4\varepsilon} = \frac{7}{48}W\left(\frac{R}{D}\right)
%                             \left(\frac{\varepsilon}{D}\right).
%\label{eq:th11}
%\end{equation}

We note here that this expression differs from that given by
Najafi and Golestanian~\cite{NG04} who reported
\begin{equation}
%U_s =
%0.7W\left(\frac{R}{D}\right)\left(\frac{\varepsilon}{D}\right)^2.
\Delta = 2.8 R \left(\frac{\varepsilon}{D}\right)^3.
\label{eq:th12}
\end{equation}
In equation (\ref{eq:th12}) the displacement, $\Delta$, is proportional
to $(\varepsilon / D)^3$. If one considers the transformation
$\varepsilon \rightarrow \chi = -\varepsilon$ and 
$D \rightarrow G = D - \varepsilon$, this corresponds to a swimmer 
undergoing exactly the same
continuous motion as that
shown in Figure \ref{fig:gol1}, only with 
the swimming stroke
beginning at the third step in the cycle. 
Thus, the swimmer must move in the same direction.
%Taking into account that
%$\Delta$ must be divided by a positive time to determine the 
%swimming speed
However, equation (\ref{eq:th12}) suggests that the swimming direction 
is reversed under this transformation, which is clearly incorrect.

In the following section, we summarize the results of
numerical simulation studies of the motion of the
linear three sphere swimmer at low Reynolds number in order
to validate and compare the use of these approaches in the study of 
swimming microstructures.

\subsection{Numerical Results}

\label{results}

\begin{figure}
\begin{center}
\includegraphics[width=3.4in,height = 3in]{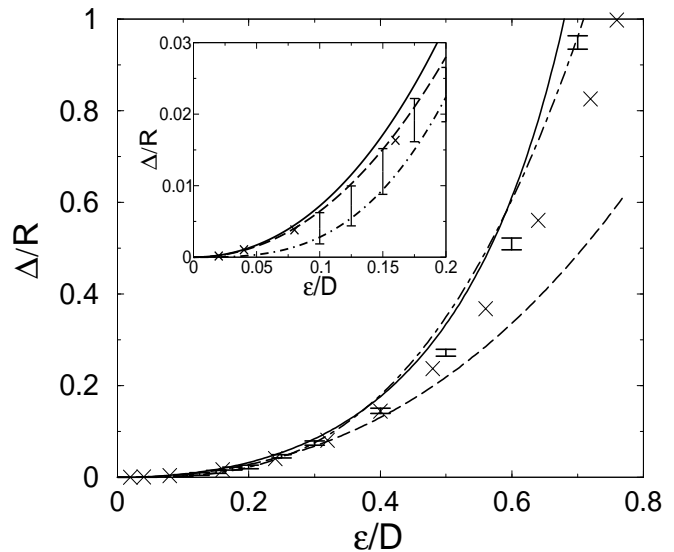}
\end{center}
\caption{
The shift per cycle of the linear three sphere swimmer, $\Delta$, 
as a function of the 
sphere displacement amplitude, $\varepsilon$. The inset shows a magnified 
view of the region of the graph below $\varepsilon/D = 0.2$.
The parameters 
$D=25$ and $R=3$ were used in the Oseen tensor and lattice Boltzmann
approaches. In the multiparticle collision dynamics simulations we
used $D=3.0 a$. The solid line was obtained by numerically solving 
the Oseen tensor equation, outlined in section \ref{OseenT}. The 
crosses mark results obtained from lattice Boltzmann simulations,
presented in section \ref{latboltsec}. The error bars show the distribution 
of results using multiparticle collision dynamics from section \ref{SRDsec}.
 The dashed line is the theoretical expression given in equation (\ref{eq:th13})
and the dot-dashed line is the expression proposed by Najafi {\em et al}.~\cite{NG04} 
reproduced in equation (\ref{eq:th12}).
}
\label{figgolelb}
\end{figure}

Figure \ref{figgolelb} gives the results for a single linear three sphere
swimmer, using each of the three methods presented in Section 
\ref{numsec}. This graph shows how the total displacement of the 
swimmer over one swimming cycle, $\Delta$, varies as a function 
of the amplitude of the stroke, $\varepsilon$. The parameters used 
were $D=25$ and $R=3$ for the Oseen tensor and lattice Boltzmann
approaches and $D = 3.0a$ for the multiparticle collision dynamics
simulations.

The solid line in Figure \ref{figgolelb} is obtained by directly
solving the Oseen tensor interaction between the spheres, 
as outlined in Section \ref{OseenT}. The dashed line shows the 
theoretical curve, equation (\ref{eq:th13}), which is correct to 
third order in $\varepsilon/D$.
These two curves converge in the limit of small $\varepsilon/D$,
as expected. 
The dot-dashed line is equation (\ref{eq:th12}), the 
expression proposed by Najafi and Golestanian~\cite{NG04}.
This appears to give good agreement for larger values of
$\varepsilon/D$. This is misleading, however, as in the 
limit of small $\varepsilon/D$ it does not converge to the theoretical
solution (this is seen most clearly within the inset), and it should 
not be valid at higher values of $\varepsilon/D$ due to the 
assumptions made in the derivation.

The lattice Boltzmann simulations 
were performed using a lattice 
of size $L_x=200$, $L_y=100$, and $L_z=100$, with periodic
boundary conditions. Initially, the 
swimmer was placed in the middle of the box, aligned parallel to 
the $x$ direction. The relaxation parameter was chosen to be 
$\tau = 1$.
 The simulations were run for one complete 
swimming cycle, which corresponded to $t_{max}=102400 \delta t$ time steps.
The maximum speed of spheres is approximated by $4\varepsilon/t_{max}$.
Using this together with $R$, which gives a characteristic length scale
for the problem, the Reynolds number can be expressed as $Re = 4 \varepsilon R/\nu %%@
t_{max}$.
%The larger $t_{max}$, the smaller the Reynolds number. 
%insert explanation about t_max here
The largest displacement used ($\varepsilon=19$) gives 
$Re = 0.013$. This was checked to be sufficiently low by running a 
limited number of simulations using $t_{max}=204800 \delta t$ time steps, 
and finding that these results agreed to within $1\%$. Furthermore, 
we checked that finite size effects were not important by performing 
simulations using a lattice of size $L_x=300$, $L_y=150$, and 
$L_z=150$, with results again agreeing within this tolerance. 

The results from the lattice Boltzmann simulations are denoted by the 
crosses in Figure \ref{figgolelb}. They agree well with the full
Oseen tensor result (solid curve) at small 
$\varepsilon/D$ and deviate as $\varepsilon/D$ gets increasingly 
large. This is because the Oseen tensor approximation that the 
spheres are far apart breaks down in this limit (the spheres
intersect each other if $\varepsilon/D > 0.76$ for the parameters
used).

Simulations using multiparticle collision dynamics are
intrinsically noisy. If we simply use a molecular dynamics algorithm
to solve the equations of motion for the swimmer and include
the particles that make up the swimmer in the solvent collision
step then, without further correction, the microstructure will rotate,
and not stay aligned along one particular axis. Although this 
behavior would be realistic for a nanoscale microstructure in a 
solvent, it does not easily allow for an accurate comparison 
of swimming speed with theory. 
To constrain the motion to one dimension, the transverse
velocities of the three particles that comprise the swimmer
were adjusted to the average velocity of the particles
after each collision step.
Changes in the arm lengths were undertaken by adding an 
extra velocity to each particle, such that 
the total momentum of the microstructure remained unchanged
during the arm length change.

For the multiparticle collision dynamics solvent we used the 
following parameters: Particle temperature
$kT = 0.005$, time step $\delta t = 0.01$, cell size $a = 1.0$, rotation angle 
$\alpha = 135^o$, average number of particles per cell $\gamma = 10$,
and particle mass $m = 10$. These result in a Reynolds number for the 
microstructure of
$\sim 10^{-5}$. These parameters both ensure a low Reynolds
number and minimize fluctuations in the solvent with 
a high Schmidt number. In our simulations, we employed a
simulation box of dimensions $30a \times 8a \times 8a$ with
periodic boundary conditions and checked for finite size
effects. 
For the swimmer we used a mass of
$5 m$ for each sphere.
%and a molecular dynamics time step of $\delta t$.
Due to the nature of the swimmer--solvent interaction in our
implementation of the multiparticle collision dynamics
algorithm, it is difficult to define the effective hydrodynamic 
radii of the spheres. However, comparison with the Oseen tensor and 
lattice Boltzmann simulation results suggests the parameters
used lead to $R/D \sim 0.12$.
The simulations were conducted for a total time of $t_{max} = 2.72
\times 10^5 \delta t$ time steps, and one period of motion took
$6.8 \times 10^3 \delta t$. The period must be sufficiently long to
allow the solvent to couple with the swimmer.
20 runs were performed for each parameter set and the results are denoted by the 
error bars in Figure \ref{figgolelb}, which are spread one standard
error on either side of the average
of the 20 runs. The results are compatible with, but much less
precise than, those obtained by the methods without intrinsic
fluctuations. 

\begin{figure}
\begin{center}
\includegraphics[width=3.2in,height = 5.5in]{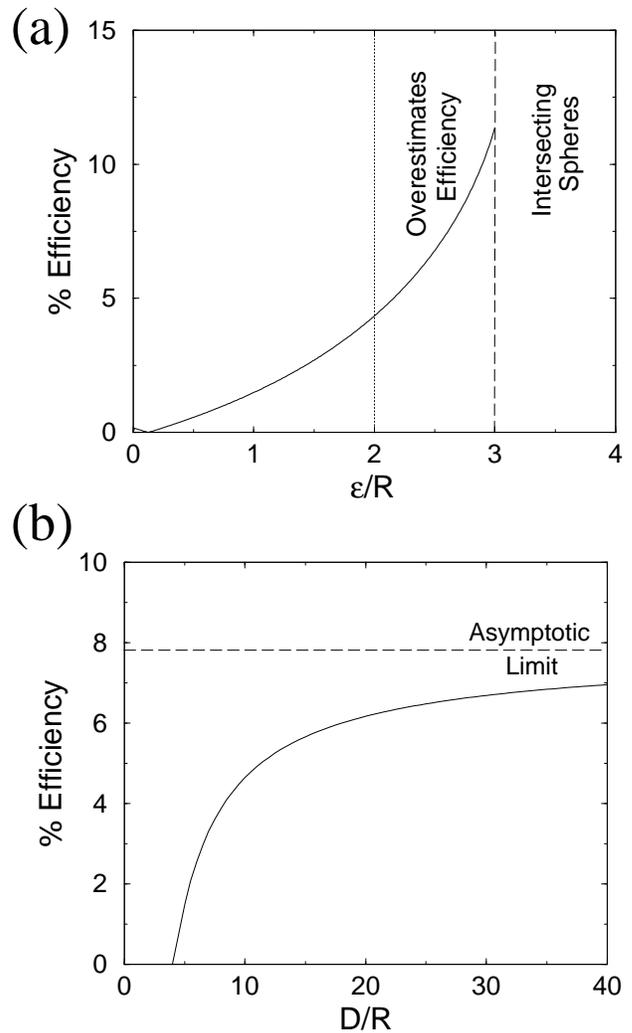}
\end{center}
\caption{ The percentage efficiency of a linear three sphere swimmer 
as a function of (a) the sphere displacement amplitude, $\varepsilon$, for fixed $D=5R$, and
(b) the maximum arm length, $D$, for fixed minimum separation $D-\varepsilon=4R$.
}
\label{effi}
\end{figure}

~\\

By changing the parameters, $D$ and $\varepsilon$, it is not 
only possible to change the swimmer displacement, $\Delta$, but also 
the efficiency of the swimmer. We define this efficiency to be the 
energy required for an external force to move the individual spheres
by the distance $\Delta$ in a time $P$ divided by the work done by the 
swimmer in performing the corresponding cyclic shape change:
\begin{eqnarray}
\text{Efficiency} = \frac{6\pi\eta N \Delta^2/P }{\sum_{i=1}^N \int_0^P \mathbf{F}_i . \mathbf{v}_i %%@
dt}.
\label{effic}
\end{eqnarray}
Using the Oseen tensor approach, the forces on the spheres, 
$\mathbf{F}_i$, are calculated through equation (\ref{F}), 
so this quantity is easily obtainable. 
The line in figure \ref{effi}(a) shows how the swimming 
efficiency is changed as a function of
$\varepsilon$, whilst keeping $D=5R$ fixed. Note that this curve 
terminates when $\varepsilon = D-2R$, since beyond this the spheres 
unphysically overlap at some point within the swimming cycle.
Furthermore, the Oseen tensor method tends to overestimate the efficiency when the spheres get close,
because it does not include the viscous dissipation resulting from lubrication effects.  
%Can we go this far with the Oseen tensor approach?
Figure \ref{effi}(a) illustrates a general feature of swimmers, namely that small amplitude motion %%@
results in inefficient swimming.
The curve in figure \ref{effi}(b) shows the efficiency against $D$, assuming a fixed mimumum sphere separation of $D-\varepsilon = 4R$. 
We find that the swimmer becomes %%@
increasingly 
efficient as $D$ increases, approaching a limit of around $8\%$.

We now summarize the relative advantages and disadvantages of the
three numerical methods. The Oseen tensor approach is particularly
advantageous for swimmers comprising small numbers of spheres because
it is computationally very fast (simulations taking only minutes
instead of days using lattice Boltzmann or multiparticle collision
dynamics). This is primarily because it is not necessary to
explicitly solve a set of fluid dynamics equations. However the
simulation time scales as $N^3$ so lattice Boltzmann becomes more
efficient for large numbers of swimmers.

The Oseen tensor formalism is also limited because it
assumes that the interacting spheres are spaced 
far apart, compared to their radii. This means, for instance, 
that it would not be appropriate 
to use this method to study the movement of a swimmer close to a wall 
or in a confined geometry. 
On the other hand, the lattice Boltzmann 
algorithm can address these problems, providing an exact 
solution to any fluid flow problem given sufficient resolution. In practice it is limited 
by computational power. To avoid spurious lattice effects, the 
sphere radius must be significantly larger than the lattice size, 
necessitating the need for large systems. Moreover, to obtain a 
sufficiently low Reynolds number, the cyclic swimming motion must 
be performed over a great number of time steps, further increasing 
the computational burden. 

Multiparticle collision dynamics is advantageous because it 
is unconditionally stable, unlike the lattice Boltzmann method, and 
this allows somewhat lower Reynolds numbers to be obtained more
easily. One can also use a molecular dynamics approach to
treat the microstructure allowing
for great flexibility in the swimmer models it is possible to consider.

Additionally, as this method inherently contains noise it will 
be appropriate for studying very small scale structures, for which 
Brownian fluctuations are important. However, if the 
fluctuations are unphysical the noise is
undesirable, necessitating long time averages.

For the remainder of this paper, we concentrate on using the Oseen 
tensor and multiparticle collision dynamics approaches to study more 
complex swimming motions.   

\section{Generalized Three Sphere Swimmers}

\label{otherthree}

\begin{figure}
\begin{center}
\includegraphics[width=3.6in,height = 3.2in]{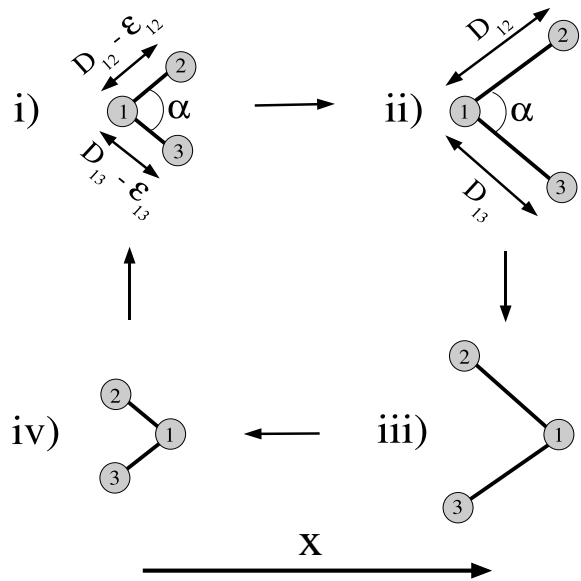}
\end{center}
\caption{
Alternative cyclic motion for a three sphere swimmer, allowing
the microstructure to translate in two dimensions. Possible extensions
to this scheme include allowing the angle $\alpha$ to change as the 
arm lengths change.
}
\label{fig3general}
\end{figure}
The swimmer described in Section \ref{1-D} is constrained to move
in one dimension along its axis.
Using the same basic elements one can design a number of other
three sphere swimmers that can move their individual 
components and centers of mass in two or three dimensions. To extend the 
original design we allow the angle between the two arms of the 
swimmer to change \cite{DB05b}. When the change in angle takes place, the 
spheres may move radially with the arm lengths 
constant, or the arm lengths may change at the same time, thus allowing 
for a number of different motions. In Figure \ref{fig3general} we show 
one of many possible alternative schemes of motion for swimmers of this 
type, which we will refer to as
generalized three sphere swimmers.
\begin{figure}
\begin{center}
\includegraphics[width=3.2in,height = 2.7in]{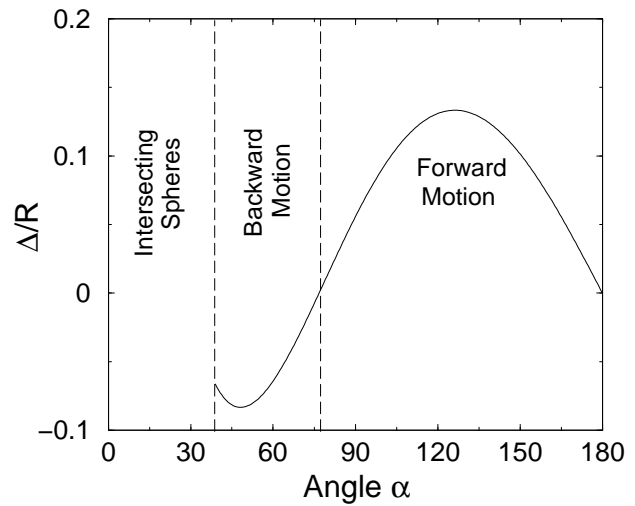}
\end{center}
\caption{
The position shift over one cycle, $\Delta$, as a function of swimmer angle $\alpha$ for %%@
the swimmer defined in Figure~\ref{fig3general}. 
Other parameters were $D = 5R$ and $\varepsilon = 2R$.
}
\label{newswim}
\end{figure}

We first concentrate
on the motion shown in Figure \ref{fig3general}, and the case where the arms 
rotate at fixed length. 
If the swimmer employs a 
symmetric motion with arm lengths $D_{12} = D_{13}$ and 
$\varepsilon_{12} = \varepsilon_{13}$, the net translation of the swimmer 
is along the {\em x}-axis (defined in the figure). Thus, the microstructure 
remains a one dimensional swimmer while its individual 
elements are moving in two dimensions.
Interestingly, the direction that the microstructure 
translates varies with $\alpha$, 
for fixed $D$ and $\varepsilon$,
as shown in Figure \ref{newswim}.
For $D = 5R$ and $\varepsilon = 2R$ the outer two spheres do not
intersect for $\alpha > 39^o$, and there 
is a transition from backward motion to forward motion at $\alpha = 77^o$.
Intuitively it is not obvious what causes this reversal of direction. 
We found that the maximum efficiency of this swimmer (with efficiency defined by equation 
(\ref{effic})), occurs at $\alpha = 138^o$. 
This corresponds to an efficiency of $1.8 \%$, which is considerably 
less than the $\sim 8 \%$ found for the linear three sphere swimmer in the 
previous section. 
It seems plausible that the linear three sphere swimmer is the most 
efficient three sphere swimmer possible.
\begin{figure}
\begin{center}
\includegraphics[width=3.2in,height = 2.7in]{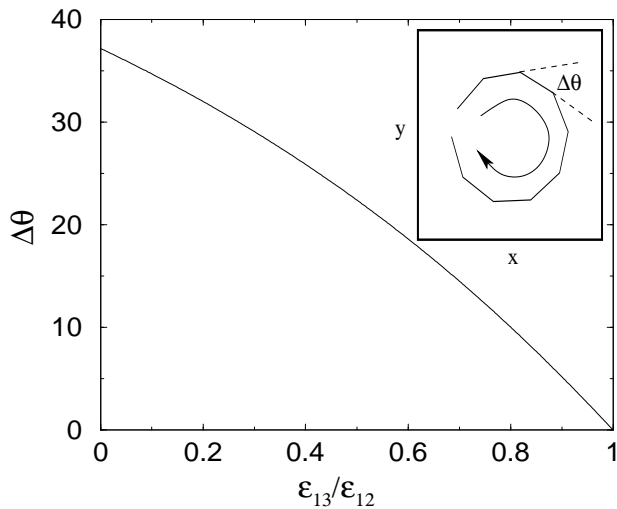}
\end{center}
\caption{The angular change in the orientation of the generalized
three sphere swimmer over one complete swimming cycle, $\Delta \theta$, 
as a function of the swimmer displacement amplitude, $\varepsilon_{13}$. 
Other parameters were fixed to be $\varepsilon_{12} = 3 R$ and $D_{12} = D_{13} = 6R$. 
The inset shows the movement of the center of mass of the swimmer at 
a fixed point in the cycle, over the course of 9 swimming cycles.}
\label{anglechange}
\end{figure}

If one instead defines an 
asymmetric motion, with $D_{12} \neq D_{13}$ and/or 
$\varepsilon_{12} \neq \varepsilon_{13}$, the microstructure will
move its center of mass in two dimensions.
The swimmer rotates and translates, as shown for
$D_{12} = D_{13} = 6$, $\varepsilon_{12} = 3$ and $\varepsilon_{13}$ 
as variable
in Figure \ref{anglechange}. As the difference between $\varepsilon_{12}$ and %%@
$\varepsilon_{13}$
becomes greater,
the angular rotation about the center of mass of the microstructure 
for each swimming cycle increases. 
Thus, it is a relatively easy step to imagine a 
manufactured device that can switch between symmetric and asymmetric 
cyclic motions,
perhaps in response to an external stimulus or experimental condition,
to enable movement in a controlled fashion in two dimensions.
A three
sphere swimmer, that can change the angle between its two arms, could 
simply adopt the efficient swimming motion of Section
\ref{1-D}, where $\alpha = 180^o$, to move in a straight line,
and then vary $\alpha$ to adopt a structure allowing it to turn.

To extend the movement of such a microstructure to three 
dimensions, the angle between the arms could be changed 
along another plane, perpendicular to the original one. 
One strategy to do this might be to employ a double-jointed structure
where the angle between the arms could be changed from 
$\alpha = 180^o$ in either of two perpendicular planes.
It is of interest that this microstructure
is the first low Reynolds number swimmer to be proposed theoretically
that can move in a controlled fashion in three dimensions, 
without employing numerous one dimensional swimming devices placed
perpendicular to each other.
%is there a reference for this??? i don't recall seeing any
%device based papers...

\section{Extended, Linear, One Dimensional Swimmers}
\label{extended}

\begin{figure}
\begin{center}
\includegraphics[width=3.0in,height = 5.0in]{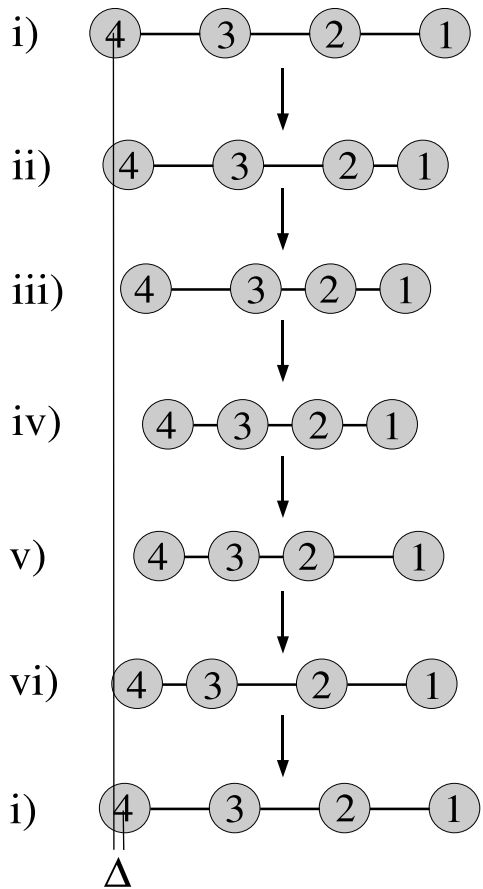}
\end{center}
\caption{
The most efficient cyclic swimming
strategy for an extended, linear, four sphere 
microstructure.
}
\label{fig4linear}
\end{figure}

To extend the linear three sphere swimmer of Section \ref{1-D} one can
simply add more spheres to the microstructure, the simplest extension
being the four sphere case. For the four sphere microstructure shown in 
Figure \ref{fig4linear}i), we analyzed through Oseen tensor based
numerical simulations all possible cyclic
motions that are made up of a discrete number of steps. At the end of each step 
the distance between neighboring spheres is either $D$ (an extended rod) or 
$D-\varepsilon$ (a contracted rod). If a rod changes length during a step then it
does so at a constant speed $W$.
In this analysis we allowed for more
than one rod length changing simultaneously. 

Out of this subset of 
possible swimmers the swimming strategy shown in Figure
\ref{fig4linear} is the most efficient. This optimal swimming strategy proceeds as
follows: Starting from the fully extended conformation, i), the
distance between spheres $1$ and $2$ is reduced to 
$D - \varepsilon$, ii). In the next two steps the distance 
between spheres $2$ and $3$ is reduced to $D - \varepsilon$, 
iii), and the distance between spheres $3$ and $4$ is reduced to 
$D - \varepsilon$, iv), the fully contracted conformation.
The microstructure then sequentially extends, first by extending the
distance between spheres $1$ and $2$ to $D$, v), then extending 
the distance between spheres $2$ and $3$ to $D$, vi), and
finally by extending the
distance between spheres $3$ and $4$ to $D$, completing the 
cycle and taking the conformation back to the original starting 
configuration i).
For the case $D = 5R$ and $\varepsilon = 2R$, the four
sphere microstructure using the optimal swimming strategy has a
swimming efficiency of 6.9 \% compared to 4.5 \% for the
three sphere swimmer using the same parameters.

\begin{figure}
\begin{center}
\includegraphics[width=3in,height = 2.5in]{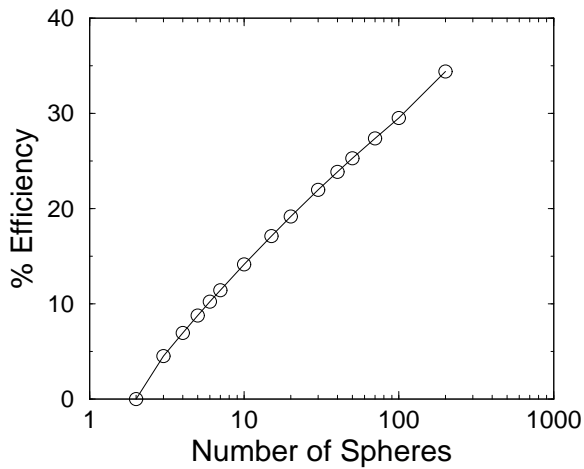}
\end{center}
\caption{Percentage efficiency against the number of spheres for a microstructure 
adopting the swimming strategy depicted in Figure \ref{fig4linear}.
}
\label{fig:beads}
\end{figure}

Extending to the five sphere case, we analyzed all possible cyclic
swimming strategies using Oseen tensor based numerical simulations, 
and found that the analogous swimming strategy
to that in Figure \ref{fig4linear} was optimal,
resulting in a swimming efficiency of 8.8 \% for the 
$D = 5R$ and $\varepsilon = 2 R$ case. We studied this optimal 
swimming strategy for microstructures with up to 200 spheres and the 
swimming efficiency as a function of the number of spheres is 
shown in Figure \ref{fig:beads}.
The curve indicates a logarithmic growth in the efficiency 
as a function of sphere number. 
This would imply that for a sufficiently large number of spheres 
the efficiency will go above one. 
Although counter intuitive, from our definition of efficiency in 
equation (\ref{effic}), this is perfectly possible, and does not violate
any physical principles. However, as the number of spheres is increased, the physical size of the spheres and rods 
would have to be made proportionately smaller, to ensure that finite Reynolds number 
effects do not become important.

\section{Filament Swimmers Using Multiparticle Collision 
Dynamics}
\label{filament}

As an example of using multiparticle collision dynamics to investigate a more complicated %%@
swimmer we consider the motion of a filament modeled as beads connected by springs and %%@
interacting through Lennard-Jones forces. The filament is driven by a sinusoidally %%@
oscillating force applied at one end. The model was motivated by the man-made microscopic %%@
swimmer of Dreyfus {\em et al.} where a red blood cell is attached to a filament %%@
consisting of superparamagnetic colloids, that are connected to each other using DNA %%@
 \cite{DB05a}. Two magnetic fields are used experimentally, one to align the filament and %%@
the other to actuate one end of it in a sinusoidal manner. This actuation results in a %%@
series of waves, originating at the tail of the filament, propagating towards the red %%@
blood cell at the head. Because the perpendicular and parallel friction coefficients of %%@
the microstructure are not equal ($\zeta_\perp / \zeta_\parallel \approx 2$), a net %%@
translation, in the opposite direction to the propagation of the wave, occurs along the %%@
alignment direction of the filament.

\begin{figure}
\begin{center}
\includegraphics[width=3.3in,height = 3.0in]{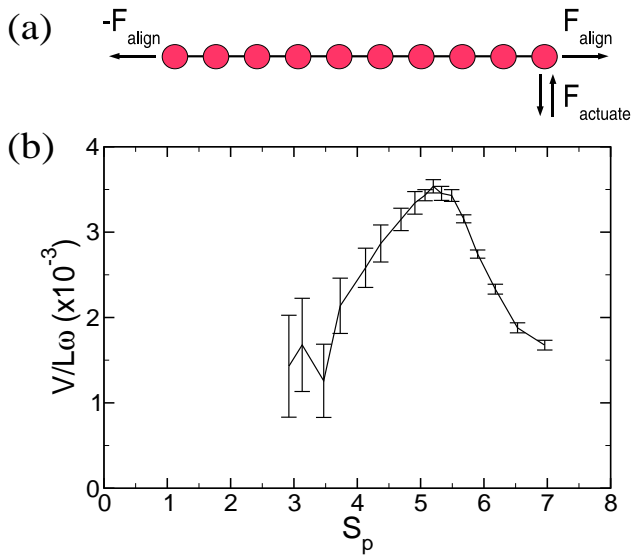}
\end{center}
\caption{(a) Model filament used in the multiparticle collision
dynamics simulations. The 
filament is composed of Lennard-Jones beads that are connected
by FENE springs. An aligning force, $F_{align}$, is applied to
both ends of the filament, and a sinusoidal force, 
$F_{actuate} = F_{max} \sin(\omega t)$, is used to actuate one
end of the microstructure, leading to wave propagation from
right to left in the diagram. 
(b) Scaled swimming velocity as a function of the sperm
number for the flexible filament. To obtain this plot, $F_{max}$ is varied
for different frequencies of actuation such that the integral of
$F_{actuate}$ over half of the period is equal.
}
\label{fig:filament}
\end{figure}

In the multiparticle collision dynamics simulation we model the filament as a number of  %%@
Lennard-Jones beads, representing the superparamagnetic colloids, connected to each  %%@
other by FENE springs, representing the DNA linkers. Instead of a magnetic %%@
field, we simply apply an equal and opposite force to each end of the filament to align %%@
it along an axis, and apply a sinusoidal actuating force, perpendicular to the aligning %%@
forces, to one end of the filament (see Figure \ref{fig:filament}(a)). 

Lowe \cite{L01} proposed a dimensionless parameter to 
characterize naturally flexible filaments, 
the `sperm number', defined as
\begin{equation}
S_p = L / \left( \frac{\kappa}{\zeta_\perp \omega} \right)^{1/4} 
\end{equation}
where $L$ is the length, $\kappa$ is the bending 
rigidity, and $\zeta_\perp$ is the perpendicular friction
coefficient for the filament, and $\omega$ is the angular
frequency of the actuation or driving. For our model filament
in a multiparticle collision dynamics solvent we calculate 
the bending rigidity from the change
in energy of the structure as a function of the curvature 
of the filament \cite{LC03}. We determine the friction coefficients
by applying a known force to each bead, in a direction perpendicular 
or parallel to the alignment, and measuring the resulting velocity of the 
microstructure (with no actuation). We can then measure how the
velocity of the microstructure depends on $S_p$ by
changing the angular frequency of the actuation.

For the multiparticle collision dynamics solvent we use 
the following parameters: 
$kT = 1.0$, $\delta t = 1.0$, $a = 1.0$, $\alpha = 120^o$, $\gamma = 5$,
and $m = 4$ 
resulting in a Reynolds number for the microstructure of
$\sim 10^{-2}$. For the microstructure we use a mass of
$4 m$ for each bead and a molecular dynamics time step of
$0.002 \delta t$. The average distance between the
centers of mass of each bead is $\approx 1.0 a$, and we
use 10 beads to represent the filament. 
The simulations are conducted over $t_{max} = 200000 \delta t$
solvent time steps, and
we average the results over 10 runs for each data point.

In Figure \ref{fig:filament} we show 
the swimming velocity of the microstructure, scaled by 
$L \omega$, as a function of $S_p$. As predicted by theory
for naturally flexible filaments \cite{WG98}, we observe a
maximum in the scaled swimming velocity of the filament between
the high (dominated by viscous friction) and low (dominated
by internal elasticity) $S_p$ regimes. This reproduces the
behavior demonstrated for the man-made swimmer 
of Dreyfus {\em et al.} \cite{DB05a}, and we also observe very similar 
scaled speeds in our simulations to those found experimentally.

To verify that the filament in the simulations 
swims through the mechanism of wave propagation, we also 
modeled a 2 bead microstructure under the influence of aligning 
and actuating forces. We find that this microstructure does not 
swim as it is impossible for a 2 bead filament to move in a 
time irreversible fashion.

\section{Conclusions}

In this paper we have described three different methods for
simulating low Reynolds number swimmers: An Oseen tensor approach, lattice Boltzmann and
multiparticle collision dynamics. In Section \ref{1-D}, each of
these methods was used to model a very simple, linear swimmer
comprising three linked spheres. Analytic results are available for
this system \cite{NG04} and hence we were able to validate the
approaches and identify the strengths and weaknesses of each
method. For swimmers made up of a small number of spheres the Oseen
tensor approach is very fast. However as the number of spheres
increases, or as multiple swimmers are considered, lattice Boltzmann
becomes more efficient. Moreover the Oseen tensor does not accurately
take into account short range hydrodynamic interactions. Lattice
Boltzmann is able to deal with spheres close to each other or to a
surface, but at the expense of an increasingly fine lattice
resolution. We found that multiparticle collision dynamics is in
general a less useful approach as the intrinsic noise tends to
dominate the results. This method will be of use when modeling
nanoscale systems where fluctuations are an intrinsic component of the
physics.

Subsequently, in Section \ref{otherthree}, we proposed a new three
sphere swimmer, which has the advantage of being able to turn, and
control its trajectory in a three dimensional manner. These ideas may
be of use in the design and fabrication of artificial microswimmers.
Section \ref{extended} extended the one dimensional three sphere
swimmer aiming to search for the most efficient swimming strategy for
a larger number of spheres. We found that the efficiency increases
logarithmically with the sphere number.  

In Section
\ref{filament}, we looked at modeling a filament swimmer
that moves due to the propagation of waves along its length. Using multiparticle
collisional dynamics we were able to reproduce the behavior
observed experimentally.

There are many directions in which it would be fruitful to pursue the
simulations. We are currently considering interactions between two or
more swimmers, and it would be of interest to consider the effect of
boundaries and obstacles on swimming behavior.
Continuum hydrodynamic theories have recently been proposed to
describe concentrated solutions of swimmers \cite{SR02,LM05}. These have led to results
very suggestive of swimming behavior but it is hard to pin down the
phenomenological parameters in the equations of motion. Simulations of
increasing numbers of swimmers are needed to try to bridge the gap
between the microscopic and continuum approaches.
 
There is enormous scope presented by more realistic
biological problems such as the chemotactic responses of bacteria or
random tumbling during a swimming cycle. Moreover simulations of this
type may be important in designing artificial microswimmers for
applications as diverse as drug delivery or mixing fluids within
microchannels.


\begin{references}
\bibitem{N2}
The dimensionless Reynolds number is defined by $Re = uL / \nu$,
where $u$ is a characteristic velocity, $L$ is a characteristic length,
and $\nu$ is the kinematic viscosity of the fluid. It is clear that 
since $\nu \sim 10^{-6}$ for a typical liquid, then on the micron scale
$Re \ll 1$.
\bibitem{N1}
This can be seen by rearranging the small Reynolds number condition to
give $R^2/\nu \ll R/u$, {\em i.e.} the time 
taken for momentum to diffuse a distance $R$ is much smaller than the time for the sphere to move that same distance.
\bibitem{N3}
Since within this framework the interior of the spheres is essentially
fluid, we lose the freedom to choose the mass density of the swimmer. 
However, for the simulations performed here, in which the 
Reynolds number is very low, inertial effects are unimportant. 
Therefore, this condition is not restrictive.
\end{references}
\end{document}